\documentclass[11pt,a4paper]{article}
\usepackage[hyperref]{acl2020}
\usepackage{times}
\usepackage{latexsym}

\usepackage{wrapfig} 
\usepackage{rotating}
\usepackage{pdflscape}

\usepackage{graphicx}
\usepackage{tabularx}
\usepackage{soul}

\usepackage{epstopdf}
\usepackage[latin1]{inputenc}

\usepackage{hyperref}
\usepackage{xstring}

\usepackage{xcolor}
\usepackage{epstopdf}

\usepackage{microtype}

\aclfinalcopy 

\title{NLP Scholar: An Interactive Visual Explorer for\\ Natural Language Processing Literature}

\author{Saif M. Mohammad \\
  National Research Council Canada \\
  \texttt{saif.mohammad@nrc-cnrc.gc.ca}}

\date{}

\begin{document}
\maketitle
\begin{abstract}
As part of the NLP Scholar project, we created a single unified dataset of NLP papers and their meta-information (including citation numbers), by extracting and aligning information from the ACL Anthology and Google Scholar.
In this paper, we describe several interconnected interactive visualizations (dashboards) that present various aspects of the data.
Clicking on an item within a visualization or entering query terms in the search boxes filters the data
in all visualizations in the dashboard. This allows users to search for papers in the area of their interest, published within specific time periods, published by specified authors, etc.
The interactive visualizations presented here, and the associated dataset of papers mapped to citations, have additional uses as well including  understanding how the field is growing (both overall and across sub-areas),
as well as
quantifying the impact of different types of papers on subsequent publications.
\end{abstract}

\section{Introduction}

NLP is a broad interdisciplinary field that draws knowledge from Computer Science, Linguistics, Information Science, Psychology, Social Sciences, 
and more.\footnote{One can make a distinction between NLP and Computational Linguistics; however, for this work we will consider them to be synonymous.} 
Over the years, scientific publications in NLP have grown in number and diversity; we now see papers published on a vast array of research questions and applications in a growing list of venues---in journals such as CL and TACL, in large conferences such as ACL and EMNLP, as well as a number of small area-focused workshops.

The ACL Anthology (AA) is a digital repository of public domain, free to access, articles on NLP.\footnote{https://www.aclweb.org/anthology/} It includes papers published in the family of ACL conferences as well as in other NLP conferences such as LREC and RANLP.
As of June 2019, it provided access to the full text and metadata for close to 50K articles published since 1965.\footnote{ACL licenses its papers with a Creative Commons Attribution 4.0 International License.} 
It is the largest single source of scientific literature on NLP. However, the meta-data does not include citation statistics.

Citation statistics are the most commonly used metrics of research impact.
They include: number of citations, average citations, h-index, relative citation ratio, and impact factor. Note, however, that the number of citations is not always a reflection of the quality or importance of a piece of work. 
Furthermore, the citation process can be abused, for example, by egregious self-citations \cite{ioannidis2019standardized}.
Nonetheless, given the immense volume of scientific literature, the relative ease with which one can track citations using services such as Google Scholar (GS), 
and given the lack of other easily applicable and effective metrics, citation analysis is an imperfect but useful window into research impact.


Google Scholar is a free web search engine for academic literature.\footnote{https://scholar.google.com}
Through it, users can access the metadata associated with an article 
such as the number of citations it has received.
Google Scholar does not provide information on how many articles are included in its database. However, scientometric researchers estimated that it included about 389 million documents in January 2018 \cite{gusenbauer2019google}---making it the world's largest source of academic information. 
Thus, it is not surprising that there is growing interest in the use of Google Scholar information to draw inferences about scholarly research  in general \cite{martin2018google,mingers2015review,orduna2014size,khabsa2014number,howland2010scholarly} 
and on scholarly impact in particular \cite{bos2019interdisciplinary,ioannidis2019standardized,ravenscroft2017measuring,bulaitis2017measuring,yogatama2011predicting,priem2010scientometrics}.


Services such as Google Scholar  and Semantic Scholar 
cover a wide variety of academic disciplines. Wile there are benefits to this, the lack of focus on NLP literature has some drawbacks as well: e.g, the potential for too many search results that include many irrelevant papers. 
For example, if one is interested in NLP papers on \textit{emotion} and \textit{privacy}, searching for them on Google Scholar
is less efficient than searching for them on a platform dedicated to NLP papers.
Further, services such as Google Scholar provide minimal interactive visualizations. NLP Scholar with its focus on AA data, is not meant to replace these tools, but act as a complementary tool for dedicated visual search of NLP literature.

ACL 2020 has a special theme asking researchers to reflect on the state of NLP.
In the spirit of that theme, and as part of a broader project on analyzing NLP Literature,
we extracted and aligned information from the ACL Anthology (AA) and Google Scholar to create a dataset of tens of thousands of NLP papers and their citations \cite{mohammad2020data,mohammad2019nlpscholar}.
In separate work, we have used the data to explores questions such as: how well cited are papers of different types (journal articles, conference papers, demo papers, etc.)?
how well cited are papers published in different time spans?
how well cited are papers from different areas of research within NLP? etc. \cite{mohammad2020citations}.
 We also explored gender gaps in Natural Language Processing research, in terms of authorship and citations \cite{mohammad2020gender}.
In this paper we describe how we built an interactive visual explorer for this unified data, which we refer to as \textit{NLP Scholar}.
 Some notable uses of NLP Scholar are listed below: 


\begin{itemize}
\item Search for relevant related work in various areas within NLP. 
\item Identify the highly cited articles on an interactive timeline. 
\item Identify past papers published in a venue of interest (such as ACL or LREC). 
\item Identify papers from the past (say ten years back) published in a venue of interest (say ACL or LREC) that have made substantial impact through citations. 
\item Examine changes in number of articles and number of citations in a chosen area of interest over time. 
\item Identify citation impact of different types of papers---e.g., short papers, shared task papers, demo papers, etc. 
\end{itemize}
\noindent Even beyond the dedicated interactive visualizer described here, the underlying data with its alignment between AA and GS has potential uses in:
\begin{itemize}
\item Creating a web browser extension that allows users of GS to look up the aligned AA  information (the full ACL BibTeX, poster, slides, access to proceedings from the same venue, etc.). 
\item Similarly, in the reverse direction, allowing access from AA to the GS information on the aligned paper. This could include number of citations, lists of papers citing the paper, etc. 
\end{itemize}
\noindent Perhaps most importantly, though, 
 NLP Scholar serves as a visual record of the state of NLP literature in terms of citations.
We note again though, that even though this work seeks to make citation metrics more accessible for ACL Anthology papers, citation metrics are not always accurate reflections of the quality, importance, or impact of individual papers.

All of the data and interactive visualizations associated with this work are freely available through the project homepage.\footnote{http://saifmohammad.com/WebPages/nlpscholar.html}

\section{Background and Related Work}

Much of the work in visualizing scientific literature has focused on showing topics of research \cite{wu2019literature,heimerl2012visual,lee2005understanding}.
There is also notable work on visualizing communities through citation networks \cite{heimerl2015citerivers,radev2016bibliometric}.

Various subsets of AA have been used in the past for a number of tasks, including:
to study citation patterns and intent \cite{radev2016bibliometric,zhu2015measuring,nanba2011classification,mohammad2009using,teufel2006automatic,aya2005citation,pham2003new},
to generate summaries of scientific articles \cite{qazvinian2013generating},
to study gender disparities in NLP \cite{schluter2018glass},
to study subtopics within NLP \cite{anderson2012towards},
and to create corpora of scientific articles \cite{mariani2018nlp4nlp,bird2008acl}.

However, none of these works 
provide an interactive visualization for users to explore NLP literature and their citations.

\section{Data}

 We now briefly describe how we extracted information from the ACL Anthology and Google Scholar. (Further details about the dataset, as well as an analysis of the volume of  research in NLP over the years,
are available in \citet{mohammad2020data}.)



\subsection{ACL Anthology Data}
The ACL Anthology provides access to its data through its website and a github repository \cite{gildea-etal-2018-acl}.\footnote{https://www.aclweb.org/anthology/\\https://github.com/acl-org/acl-anthology}
We extracted paper title, names of authors, year of publication, and venue of publication from the repository.\footnote{Multiple authors can have the same name and the same authors may use multiple variants of their names in papers. 
The AA volunteer team handles such ambiguities using both semi-automatic and manual approaches (fixing some instances on a case-by-case basis). 
Additionally, the AA repository includes a file that has canonical forms of author names. Authors can provide AA with their aliases, change-of-name information, and preferred canonical name, which is then eventually recorded in the canonical-name file.}


As of June 2019, AA had $\sim$50K entries; however, this includes forewords, schedules, etc.\@
that are not truly research publications. 
After discarding them we are left with a set of 44,895 papers.


\subsection{Google Scholar Data}


Google Scholar does not provide an API to extract information about the papers. This is likely because of its agreement with publishing companies that have scientific literature behind paywalls \cite{martin2018google}.  We extracted citation information from Google Scholar profiles of authors who published at least three papers in the ACL Anthology. 
(This is explicitly allowed by GS's robots exclusion standard. This is also how past work has studied Google Scholar \cite{khabsa2014number,orduna2014size,martin2018google}.)
This yielded citation information for 1.1 million papers in total. We will refer to this dataset as 
\textit{GS-NLP}. 
Note that GS-NLP includes citation counts not just for NLP papers, but also for non-NLP papers published by the authors. 

GS-NLP includes 32,985 
of the 44,895 papers in AA (about 74\%). We will refer to this subset of the ACL Anthology papers as AA$'$. The citation analyses presented in this paper are on AA$'$.
(Future work will explore visualizations on GS-NLP.)

Entries across AA and GS are aligned by matching the paper title, year of publication, and first author last name.\footnote{There were marked variations in how the same venue was described in the meta-information across AA and GS; thus, venue information was not used for alignment.}

\section{Building an Interactive Visualization to Explore Scientific Literature}


We now describe how we created an interactive visualization---NLP Scholar---that allows one to visually explore the data from the ACL Anthology along with citation information from Google Scholar. We first created a relational database (involving multiple tables) that stores the AA and GS data (\S\ref{sec:db}). We then loaded the database in Tableau---an interactive data visualization software---to build the visualizations (\S\ref{sec:tableau-viz}).\footnote{Tableau: https://www.tableau.com\\ Even though there are paid versions of Tableau, the visualizations built with Tableau can be freely shared with others on the world wide web. Users do not require any special software to interact with these visualization on the web.}

\subsection{NLP Scholar Relational Database}
\label{sec:db}

Data from AA and GS is stored in four tables (tsv files): papers, authors, title-unigrams, and title-bigrams.
They contain the following information:\\[-12pt]

\noindent {\bf papers:} Each row corresponds to a unique paper. The columns include: paper title, year of publication, list of authors, venue of publication, number of citations at the time of data collection (June 2019),  NLP Scholar paper id, 
ACL paper id, 
and some other meta-data associated with the paper. 

The \textit{NLP Scholar paper id} is a concatenation of the paper title, year of publication, and first author last name. (This id was also used to align entries across AA and GS).\\[-12pt]

\noindent {\bf authors:} Each row corresponds to a paper--author combination. The columns include: NLP Scholar paper id, author first name, and author last name. A paper with three authors contributes three rows to the table (all three have the same paper id, but different author names).\\[-12pt]

\noindent {\bf title-unigrams:} Each row corresponds to a paper title and unigram combination. The columns include: NLP Scholar paper id and paper title unigram (a word that occurs in the title of the paper). A paper with five unique words in the title contributes five rows to the table (all five have the same paper id, but different words).\\[-12pt]

\noindent {\bf title-bigrams:} Each row corresponds to a paper title and bigram combination. The columns include: NLP Scholar paper id and paper title bigram (a two-word sequence that occurs in the title of the paper). A paper with four unique bigrams in the title contributes four rows to the table (all four have the same paper id, but different bigrams).\\[-12pt]

Once the tables are loaded in Tableau, the following pairs of tables are each joined (inner join) using the NLP Scholar paper id:\footnote{An inner join selects all rows from both participating tables whose join column values match across the two tables.} 
papers--authors, papers--title-unigrams, and papers--title-bigrams.

\begin{figure*}[t!]
 \begin{center}
 	\includegraphics[width=2.08\columnwidth]{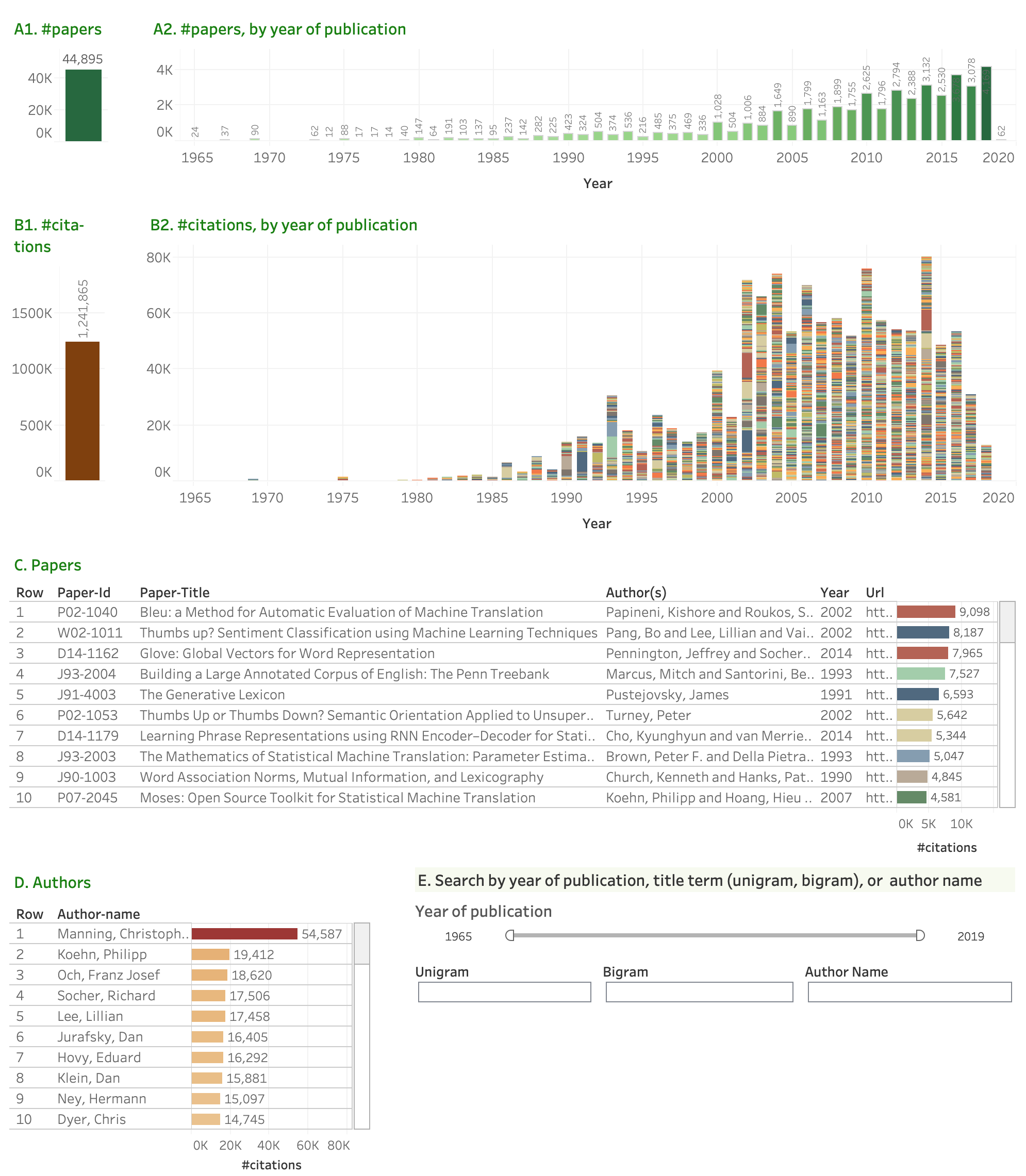}
 	\caption{A screenshot of NLP Scholar's principle dashboard.}
 	\label{fig:main}
 \end{center}
 \vspace*{4mm}
\end{figure*}

\subsection{NLP Scholar Interactive Visualization}
\label{sec:tableau-viz}

We developed multiple  visualizations to explore various aspects of the data. 
We group and connect several individual visualizations in dashboards that allow one to explore several aspects of the data together.  Clicking on data attributes such as year of publication or venue of publication in one visualization, filters the data in all visualizations within a dashboard to show only the relevant data. 

Figure \ref{fig:main} shows a screenshot of the main dashboard. At the top are the number of papers---total (A1) and by year of publication (A2). This allows one to see the growth/decline of the papers over the years. 

Below it, we see the number of citations---total (B1) and by year of publication (B2). For a given year, the bar is partitioned into segments corresponding to individual papers. Each segment (paper) has a height that is proportional to the number of citations it has received and assigned a colour at random. This allows one to quickly identify high-citation papers.\footnote{Note that since the number of colours is smaller than the number of papers, multiple papers may have the same color; however, the probability of adjacent papers receiving the same colour is small---even then, the system will provide visual clues distinguishing each segment when hovering over the area.}

Hovering over individual papers in B2 pops open an  information box showing the paper title, authors, year of publication, publication venue, and \#citations. Figure 6 in the Appendix shows a blow up of B2 along with examples of the hover information box.
Similarly, hovering over other parts of the dashboard shows corresponding information. (This is especially helpful, when parts of the text are truncated or otherwise not visible due to space constraints.)

Further below, we see lists of papers (C) and authors (D)---both are ordered by number of citations.
Search boxes in the bottom right (E) allow searching for papers that have particular terms in the title or searching for papers by author name. One can also restrict the search to a span of years using the slider.

\begin{figure*}[t!]
 \begin{center}
 	\includegraphics[width=2\columnwidth]{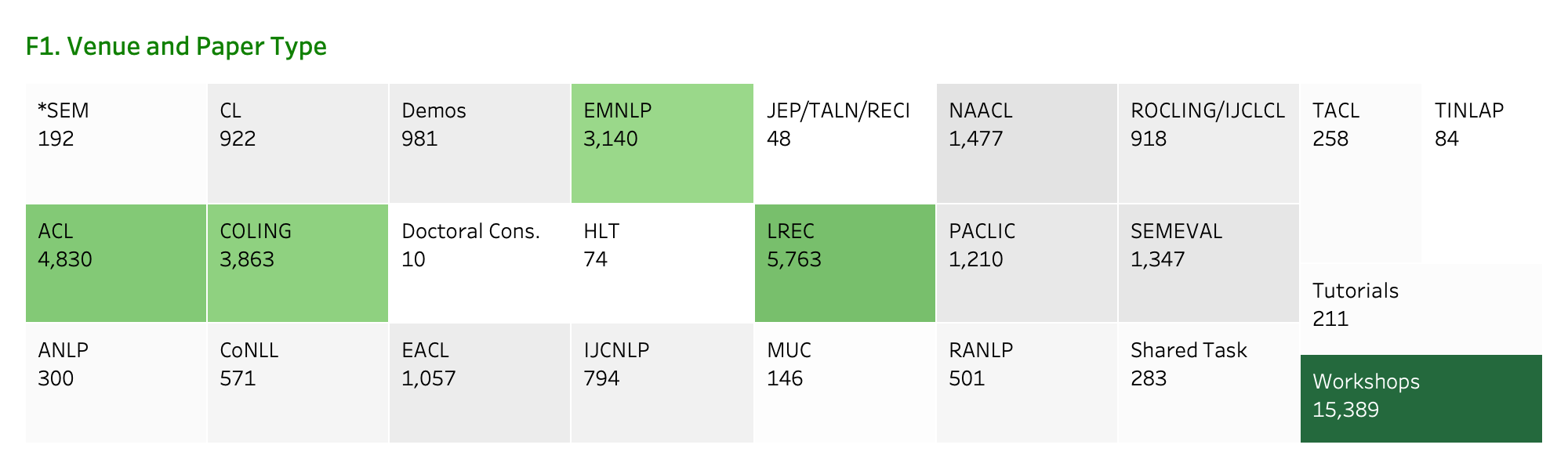}
 	\caption{A treemap of popular NLP venues and paper types. Darker shades of green: higher volumes of papers.}
 	\label{fig:venues}
 \end{center}
\end{figure*}
\begin{figure*}[t!]
 \begin{center}
 	\includegraphics[width=2\columnwidth]{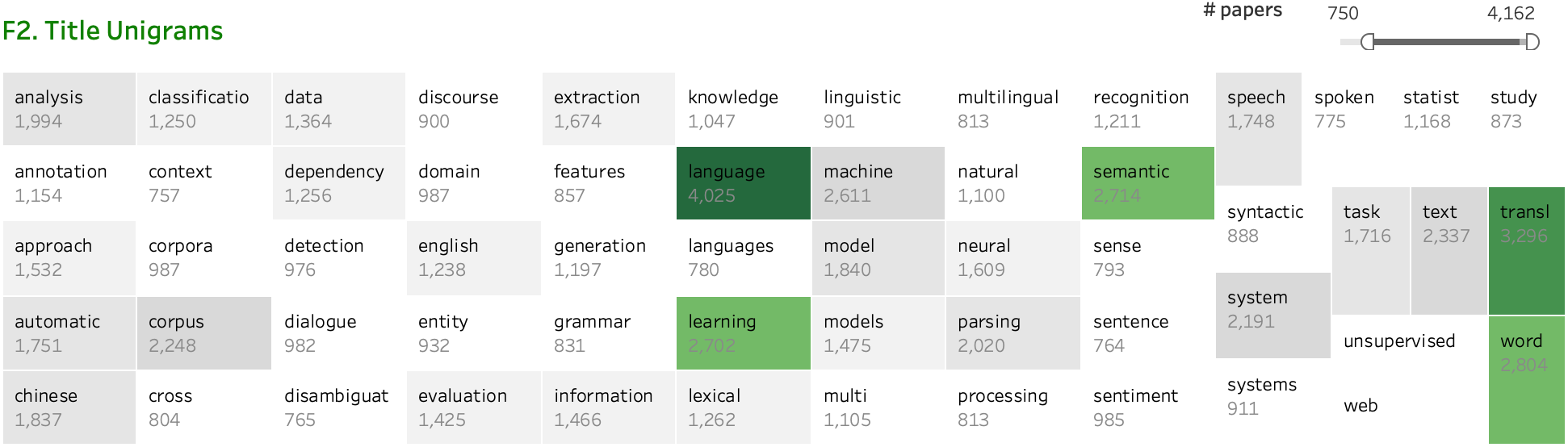}
 	\caption{A treemap of the most common unigrams in paper titles. Darker shades of green: higher frequencies.}
 	\label{fig:unigrams}
 \end{center}
\end{figure*}
\begin{figure*}[t!]
 \begin{center}
 	\includegraphics[width=2\columnwidth]{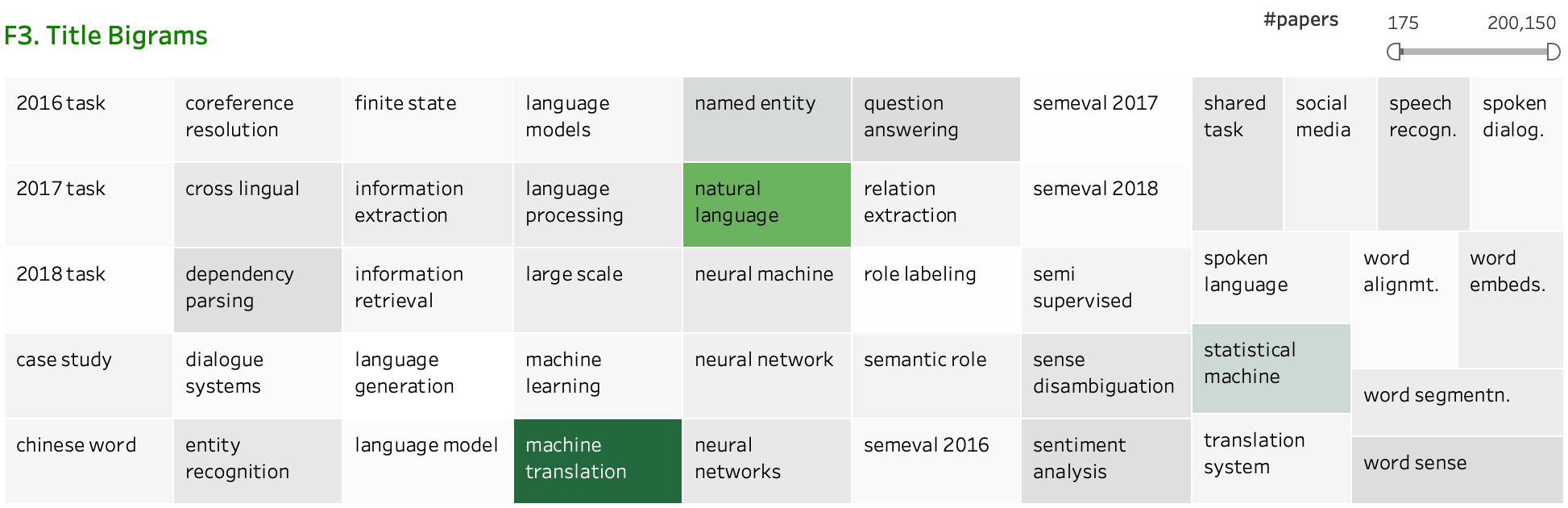}
 	\caption{A treemap of the most common bigrams in paper titles. Darker shades of green: higher frequencies.}
 	\label{fig:bigrams}
 \end{center}
\end{figure*}
\begin{figure*}[t!]
 \begin{center}
 	\includegraphics[width=2\columnwidth]{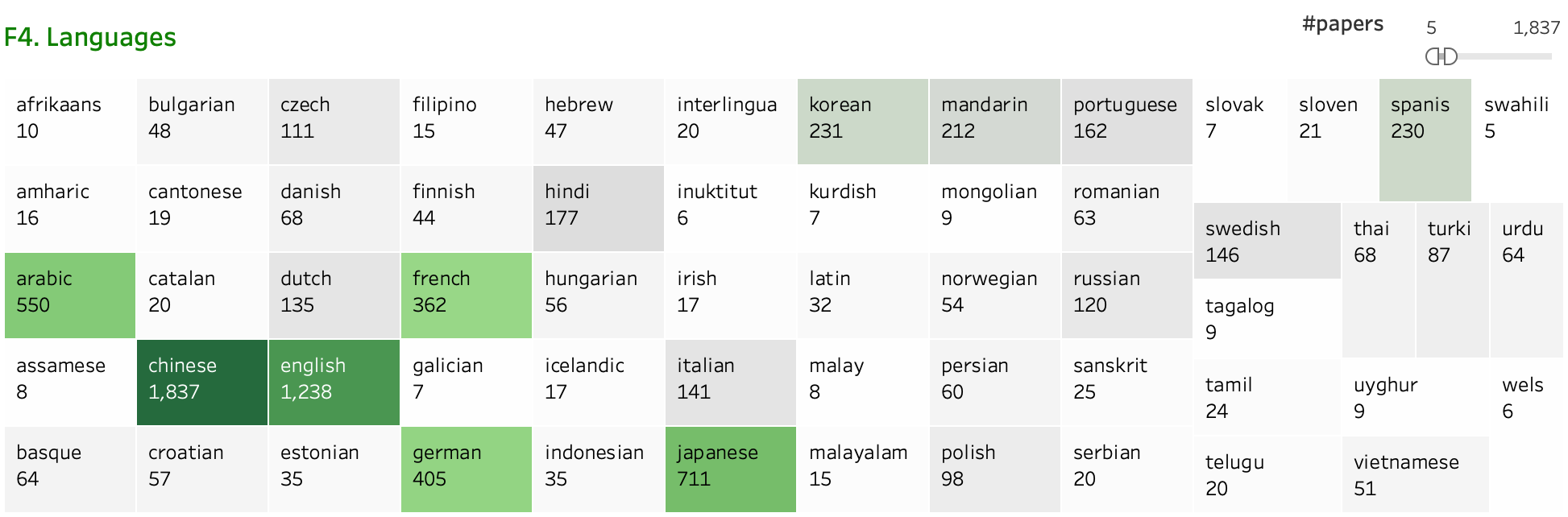}
 	\caption{A treemap of the most common language terms in titles. Darker shades of green: higher frequencies.}
 	\label{fig:languages}
 \end{center}
\end{figure*}

Four other dashboards are also created that have the same five elements as the main dashboard (A through E), and additionally include a six element F to provide a focused search facility.
This sixth element is a treemap that shows the most common: venues and paper types (F1), title unigrams (F2), title bigrams (F3), or language mentions in the title (F4). (We only show one of the four treemaps at a time to prevent overwhelming the user.)
The treemaps are shown in Figures \ref{fig:venues} to \ref{fig:languages}, respectively.

\section{Data Explorations with NLP Scholar}
\label{sec:explore}

\noindent Figure \ref{fig:main} A1 shows that the dataset includes 44,895 papers. 
A2 shows that the volume of papers published was considerably lower in the early years (1965 to 1989);
there was a spurt in the 1990s; and substantial numbers since the year 2000. Also, note that the number of publications is considerably higher in alternate years. This is due to certain biennial conferences. Since 1998 the largest of such conferences has been LREC (In 2018 alone LREC had over 700 main conferences papers and additional papers from its 29 workshops). COLING, another biennial conference (also occurring in the even years) has about 45\% of the number of main conference papers as LREC.

B1 shows that AA$'$ papers have received $\sim$1.2 million citations (as of June 2019). The timeline graph in B2 shows that, with time, not only have the number of papers grown, but also the number of high-citation papers. We see a marked jump in the 1990s over the previous decades, but the 2000s are the most notable in terms of the high number of citations. The 2010s papers will likely surpass the 2000s papers in the years to come.

The most cited papers list (C) shows influential papers from machine translation, sentiment analysis, word embeddings, syntax, and semantics.

Among the authors (D), observe that Christopher Manning has 
not only received the most number of citations, 
he has also received almost three times as many citations as the next person in the list.

\noindent {\bf Search:} NLP Scholar allows for search in a number of ways.
Suppose we are interested in the topic of sentiment analysis. Then we can enter the relevant keywords in the search box: \textit{sentiment, valence, emotion, emotions, affect,} etc.
Then the visualizations are filtered to present details of only those papers that have at least one of these keywords in the title. (Future work will allow for search in the abstract and the whole text.)

Figure 7 in the Appendix shows the filtered result. The system identified 1,481 papers that each have at least one of the query terms in the title. They have received more than 85K citations. 
The citations timeline (B2 in Figure 7) shows that there were just a few scattered papers in early years (1987--2000) that received a small number of citations.
However, two papers in 2002 received a massive number of citations, and likely led to the substantially increased interest in the field.  The number of papers has steadily increased since 2002, with close to 250 papers in 2018, showing that the area continues to enjoy considerable attention. 

One 
can also fine tune the search as desired. Say we are interested not in the broad area of sentiment analysis, but specifically in the work on emotions and affect. Then they can enter only emotion- and affect-related keywords. A disadvantage of using terms for search is that some terms are ambiguous and they can pull in irrelevant articles; also if a paper is about the topic of interest but its title does not have one of the standard keywords associated with the topic, then it might be left out. 
That said, if one does come across a paper that has the query term but is not in the topic of interest, they can right click and exclude that paper from the visualization; and as mentioned before, future work will allow for searches in the abstract and full text as well.
We are also currently working on clustering papers using the words in the articles as features.\footnote{Note that clustering approaches also have limitations, such as differing results depending on the parameters used.}


Below are some more examples of interactions with NLP Scholar (Figures are in the Appendix after references):\\[-18pt]
\begin{itemize}
\item Figure 8 shows the state of the visualization when one clicks the year 2016 in A1. 
\item Figures 9 and 10 show examples of author search by clicking on the authors list (D) (\textit{Christopher Manning} and \textit{Lillian Lee}). 
\item Figures 11 and 12 show the dashboard 
when one clicks on the 
Venue and Paper Type treemap (F1): \textit{ACL main conference papers} and \textit{Workshop papers}, respectively. 
\item Figures 13, 14 and 15 in the Appendix also show examples of search for the terms 
\textit{parsing, statistical} and \textit{neural}, respectively (accessed by clicking on the title unigrams treemap (F2)). 
\item Figures 16, 17, and 18 show the dashboard when one clicks on the 
Title Bigrams treemap (F3): \textit{machine translation, question answering,} and \textit{word embeddings}, respectively. 
\item Figures 19 and 20 show the dashboard when one clicks on the 
Languages treemap (F4): \textit{Chinese} and \textit{Swahili}, respectively. 
\end{itemize}
\noindent Once the system goes live, we hope to collect further usage scenarios from the users at large.

For this work, we chose not to stem the terms in the titles before applying the search. This is because in some search scenarios,
it is beneficial to distinguish the different morphological forms of a word. For example,
papers with \textit{emotions} in the titles are more likely to be dealing with multiple emotions
than papers with the term {\it emotion}. When such distinctions do not need to be made, it is easy for users to include  morphological variants as additional query terms.

\section{Conclusions and Future Work}

We presented NLP Scholar---an interactive visual explorer for the ACL Anthology.
Notably, the tool also has access to citation information from Google Scholar.
It includes several interconnected interactive visualizations (dashboards) that allow users to quickly and efficiently search for relevant related work by
clicking on items within a visualization or through search boxes.
All of the data and interactive visualizations associated with this work are freely available through the project homepage.\footnote{http://saifmohammad.com/WebPages/nlpscholar.html} 

Future work will provide additional functionalities such as search within abstracts and whole texts, document clustering, and automatically identifying related papers. 
We see NLP Scholar, with its dedicated visual search capabilities for NLP papers, as a useful complementary tool to existing resources such as Google Scholar. 
We also note that the approach presented here is not required to be applied only to the ACL Anthology or NLP papers;
it can be used to display papers from other sources too such as pre-print archives and anthologies of papers
from other fields of study.




\section*{Acknowledgments}

This work was possible due to the helpful discussion and encouragement from a number of awesome people including: Dan Jurafsky, Tara Small, Michael Strube, Cyril Goutte, Eric Joanis, Matt Post, Torsten Zesch, Ellen Riloff, Iryna Gurevych, Rebecca Knowles, Isar Nejadgholi, and Peter Turney. Also, a big thanks to the ACL Anthology and Google Scholar Teams for creating and maintaining wonderful resources.

\bibliography{ACL2020-NLP-Scholar-Demo}

\begin{thebibliography}{33}
\expandafter\ifx\csname natexlab\endcsname\relax\def\natexlab#1{#1}\fi

\bibitem[{Anderson et~al.(2012)Anderson, McFarland, and
  Jurafsky}]{anderson2012towards}
Ashton Anderson, Dan McFarland, and Dan Jurafsky. 2012.
\newblock Towards a computational history of the acl: 1980-2008.
\newblock In \emph{Proceedings of the ACL-2012 Special Workshop on
  Rediscovering 50 Years of Discoveries}, pages 13--21. Association for
  Computational Linguistics.

\bibitem[{Aya et~al.(2005)Aya, Lagoze, and Joachims}]{aya2005citation}
Selcuk Aya, Carl Lagoze, and Thorsten Joachims. 2005.
\newblock Citation classification and its applications.
\newblock In \emph{Knowledge Management: Nurturing Culture, Innovation, and
  Technology}, pages 287--298. World Scientific.

\bibitem[{Bird et~al.(2008)Bird, Dale, Dorr, Gibson, Joseph, Kan, Lee, Powley,
  Radev, and Tan}]{bird2008acl}
Steven Bird, Robert Dale, Bonnie Dorr, Bryan Gibson, Mark Joseph, Min-Yen Kan,
  Dongwon Lee, Brett Powley, Dragomir Radev, and Yee~Fan Tan. 2008.
\newblock The {ACL} anthology reference corpus: A reference dataset for
  bibliographic research in computational linguistics.
\newblock In \emph{Proceedings of the Sixth International Conference on
  Language Resources and Evaluation ({LREC}'08)}, Marrakech, Morocco. European
  Language Resources Association (ELRA).

\bibitem[{Bos and Nitza(2019)}]{bos2019interdisciplinary}
Arthur~R Bos and Sandrine Nitza. 2019.
\newblock Interdisciplinary comparison of scientific impact of publications
  using the citation-ratio.
\newblock \emph{Data Science Journal}, 18(1).

\bibitem[{Bulaitis(2017)}]{bulaitis2017measuring}
Zoe Bulaitis. 2017.
\newblock Measuring impact in the humanities: Learning from accountability and
  economics in a contemporary history of cultural value.
\newblock \emph{Palgrave Communications}, 3(1):7.

\bibitem[{Gildea et~al.(2018)Gildea, Kan, Madnani, Teichmann, and
  Villalba}]{gildea-etal-2018-acl}
Daniel Gildea, Min-Yen Kan, Nitin Madnani, Christoph Teichmann, and Mart{\'\i}n
  Villalba. 2018.
\newblock \href {https://doi.org/10.18653/v1/W18-2504} {The {ACL} anthology:
  Current state and future directions}.
\newblock In \emph{Proceedings of Workshop for {NLP} Open Source Software
  ({NLP}-{OSS})}, pages 23--28, Melbourne, Australia.

\bibitem[{Gusenbauer(2019)}]{gusenbauer2019google}
Michael Gusenbauer. 2019.
\newblock Google scholar to overshadow them all? comparing the sizes of 12
  academic search engines and bibliographic databases.
\newblock \emph{Scientometrics}, 118(1):177--214.

\bibitem[{Heimerl et~al.(2015)Heimerl, Han, Koch, and
  Ertl}]{heimerl2015citerivers}
Florian Heimerl, Qi~Han, Steffen Koch, and Thomas Ertl. 2015.
\newblock Citerivers: Visual analytics of citation patterns.
\newblock \emph{IEEE transactions on visualization and computer graphics},
  22(1):190--199.

\bibitem[{Heimerl et~al.(2012)Heimerl, Koch, Bosch, and
  Ertl}]{heimerl2012visual}
Florian Heimerl, Steffen Koch, Harald Bosch, and Thomas Ertl. 2012.
\newblock Visual classifier training for text document retrieval.
\newblock \emph{IEEE Transactions on Visualization and Computer Graphics},
  18(12):2839--2848.

\bibitem[{Howland et~al.(2009)Howland, Wright, Boughan, and
  Roberts}]{howland2010scholarly}
Jared~L. Howland, Thomas~C. Wright, Rebecca~A. Boughan, and Brian~C. Roberts.
  2009.
\newblock How scholarly is google scholar? a comparison to library databases.
\newblock \emph{College \& Research Libraries}, 70(3).

\bibitem[{Ioannidis et~al.(2019)Ioannidis, Baas, Klavans, and
  Boyack}]{ioannidis2019standardized}
John~PA Ioannidis, Jeroen Baas, Richard Klavans, and Kevin~W Boyack. 2019.
\newblock A standardized citation metrics author database annotated for
  scientific field.
\newblock \emph{PLoS biology}, 17(8):e3000384.

\bibitem[{Khabsa and Giles(2014)}]{khabsa2014number}
Madian Khabsa and C~Lee Giles. 2014.
\newblock The number of scholarly documents on the public web.
\newblock \emph{PloS one}, 9(5):e93949.

\bibitem[{Lee et~al.(2005)Lee, Czerwinski, Robertson, and
  Bederson}]{lee2005understanding}
Bongshin Lee, Mary Czerwinski, George Robertson, and Benjamin~B Bederson. 2005.
\newblock Understanding research trends in conferences using paperlens.
\newblock In \emph{CHI'05 extended abstracts on Human factors in computing
  systems}, pages 1969--1972.

\bibitem[{Mariani et~al.(2018)Mariani, Francopoulo, and
  Paroubek}]{mariani2018nlp4nlp}
Joseph Mariani, Gil Francopoulo, and Patrick Paroubek. 2018.
\newblock The nlp4nlp corpus (i): 50 years of publication, collaboration and
  citation in speech and language processing.
\newblock \emph{Frontiers in Research Metrics and Analytics}, 3:36.

\bibitem[{Mart{\'\i}n-Mart{\'\i}n et~al.(2018)Mart{\'\i}n-Mart{\'\i}n,
  Orduna-Malea, Thelwall, and L{\'o}pez-C{\'o}zar}]{martin2018google}
Alberto Mart{\'\i}n-Mart{\'\i}n, Enrique Orduna-Malea, Mike Thelwall, and
  Emilio~Delgado L{\'o}pez-C{\'o}zar. 2018.
\newblock Google scholar, web of science, and scopus: A systematic comparison
  of citations in 252 subject categories.
\newblock \emph{Journal of Informetrics}, 12(4):1160--1177.

\bibitem[{Mingers and Leydesdorff(2015)}]{mingers2015review}
John Mingers and Loet Leydesdorff. 2015.
\newblock A review of theory and practice in scientometrics.
\newblock \emph{European journal of operational research}, 246(1):1--19.

\bibitem[{Mohammad et~al.(2009)Mohammad, Dorr, Egan, Hassan, Muthukrishan,
  Qazvinian, Radev, and Zajic}]{mohammad2009using}
Saif Mohammad, Bonnie Dorr, Melissa Egan, Ahmed Hassan, Pradeep Muthukrishan,
  Vahed Qazvinian, Dragomir Radev, and David Zajic. 2009.
\newblock Using citations to generate surveys of scientific paradigms.
\newblock In \emph{Proceedings of human language technologies: The 2009 annual
  conference of the North American chapter of the association for computational
  linguistics}, pages 584--592.

\bibitem[{Mohammad(2019)}]{mohammad2019nlpscholar}
Saif~M. Mohammad. 2019.
\newblock The state of nlp literature: A diachronic analysis of the acl
  anthology.
\newblock \emph{arXiv preprint arXiv:1911.03562}.

\bibitem[{Mohammad(2020{\natexlab{a}})}]{mohammad2020citations}
Saif~M. Mohammad. 2020{\natexlab{a}}.
\newblock Examining citations of natural language processing literature.
\newblock In \emph{Proceedings of the 2020 Annual Conference of the Association
  for Computational Linguistics}, Seattle, USA.

\bibitem[{Mohammad(2020{\natexlab{b}})}]{mohammad2020gender}
Saif~M. Mohammad. 2020{\natexlab{b}}.
\newblock Gender gap in natural language processing research: Disparities in
  authorship and citations.
\newblock In \emph{Proceedings of the 2020 Annual Conference of the Association
  for Computational Linguistics}, Seattle, USA.

\bibitem[{Mohammad(2020{\natexlab{c}})}]{mohammad2020data}
Saif~M. Mohammad. 2020{\natexlab{c}}.
\newblock Nlp scholar: A dataset for examining the state of nlp research.
\newblock In \emph{Proceedings of the 12th Language Resources and Evaluation
  Conference (LREC-2020)}, Marseille, France.

\bibitem[{Nanba et~al.(2011)Nanba, Kando, and
  Okumura}]{nanba2011classification}
Hidetsugu Nanba, Noriko Kando, and Manabu Okumura. 2011.
\newblock Classification of research papers using citation links and citation
  types: Towards automatic review article generation.
\newblock \emph{Advances in Classification Research Online}, 11(1):117--134.

\bibitem[{Ordu{\~n}a-Malea et~al.(2014)Ordu{\~n}a-Malea, Ayll{\'o}n,
  Mart{\'\i}n-Mart{\'\i}n, and L{\'o}pez-C{\'o}zar}]{orduna2014size}
Enrique Ordu{\~n}a-Malea, Juan~Manuel Ayll{\'o}n, Alberto
  Mart{\'\i}n-Mart{\'\i}n, and Emilio~Delgado L{\'o}pez-C{\'o}zar. 2014.
\newblock About the size of google scholar: playing the numbers.
\newblock \emph{arXiv preprint arXiv:1407.6239}.

\bibitem[{Pham and Hoffmann(2003)}]{pham2003new}
Son~Bao Pham and Achim Hoffmann. 2003.
\newblock A new approach for scientific citation classification using cue
  phrases.
\newblock In \emph{Australasian Joint Conference on Artificial Intelligence},
  pages 759--771. Springer.

\bibitem[{Priem and Hemminger(2010)}]{priem2010scientometrics}
Jason Priem and Bradely~H Hemminger. 2010.
\newblock Scientometrics 2.0: New metrics of scholarly impact on the social
  web.
\newblock \emph{First monday}, 15(7).

\bibitem[{Qazvinian et~al.(2013)Qazvinian, Radev, Mohammad, Dorr, Zajic,
  Whidby, and Moon}]{qazvinian2013generating}
Vahed Qazvinian, Dragomir~R Radev, Saif~M Mohammad, Bonnie Dorr, David Zajic,
  Michael Whidby, and Taesun Moon. 2013.
\newblock Generating extractive summaries of scientific paradigms.
\newblock \emph{Journal of Artificial Intelligence Research}, 46:165--201.

\bibitem[{Radev et~al.(2016)Radev, Joseph, Gibson, and
  Muthukrishnan}]{radev2016bibliometric}
Dragomir~R Radev, Mark~Thomas Joseph, Bryan Gibson, and Pradeep Muthukrishnan.
  2016.
\newblock A bibliometric and network analysis of the field of computational
  linguistics.
\newblock \emph{Journal of the Association for Information Science and
  Technology}, 67(3):683--706.

\bibitem[{Ravenscroft et~al.(2017)Ravenscroft, Liakata, Clare, and
  Duma}]{ravenscroft2017measuring}
James Ravenscroft, Maria Liakata, Amanda Clare, and Daniel Duma. 2017.
\newblock Measuring scientific impact beyond academia: An assessment of
  existing impact metrics and proposed improvements.
\newblock \emph{PloS one}, 12(3):e0173152.

\bibitem[{Schluter(2018)}]{schluter2018glass}
Natalie Schluter. 2018.
\newblock The glass ceiling in {NLP}.
\newblock In \emph{Proceedings of the 2018 Conference on Empirical Methods in
  Natural Language Processing}, pages 2793--2798.

\bibitem[{Teufel et~al.(2006)Teufel, Siddharthan, and
  Tidhar}]{teufel2006automatic}
Simone Teufel, Advaith Siddharthan, and Dan Tidhar. 2006.
\newblock Automatic classification of citation function.
\newblock In \emph{Proceedings of the 2006 Conference on Empirical Methods in
  Natural Language Processing}, pages 103--110.

\bibitem[{Wu et~al.(2019)Wu, Zhao, Parvinzamir, Ersotelos, Wei, and
  Dong}]{wu2019literature}
Shaopeng Wu, Youbing Zhao, Farzad Parvinzamir, Nikolaos~Th Ersotelos, Hui Wei,
  and Feng Dong. 2019.
\newblock Literature explorer: effective retrieval of scientific documents
  through nonparametric thematic topic detection.
\newblock \emph{The Visual Computer}, pages 1--18.

\bibitem[{Yogatama et~al.(2011)Yogatama, Heilman, O'Connor, Dyer, Routledge,
  and Smith}]{yogatama2011predicting}
Dani Yogatama, Michael Heilman, Brendan O'Connor, Chris Dyer, Bryan~R
  Routledge, and Noah~A Smith. 2011.
\newblock Predicting a scientific community's response to an article.
\newblock In \emph{Proceedings of the 2011 Conference on Empirical Methods in
  Natural Language Processing}, pages 594--604.

\bibitem[{Zhu et~al.(2015)Zhu, Turney, Lemire, and Vellino}]{zhu2015measuring}
Xiaodan Zhu, Peter Turney, Daniel Lemire, and Andr{\'e} Vellino. 2015.
\newblock Measuring academic influence: Not all citations are equal.
\newblock \emph{Journal of the Association for Information Science and
  Technology}, 66(2):408--427.

\end{thebibliography}
\bibliographystyle{acl_natbib}

\appendix

\section{Appendix}

Figures 6 through 20 (in the pages ahead) show example interactions with NLP Scholar that were discussed in Section \ref{sec:explore}.


\begin{sidewaysfigure*}[t]
 \begin{center}
 	\includegraphics[width=\columnwidth]{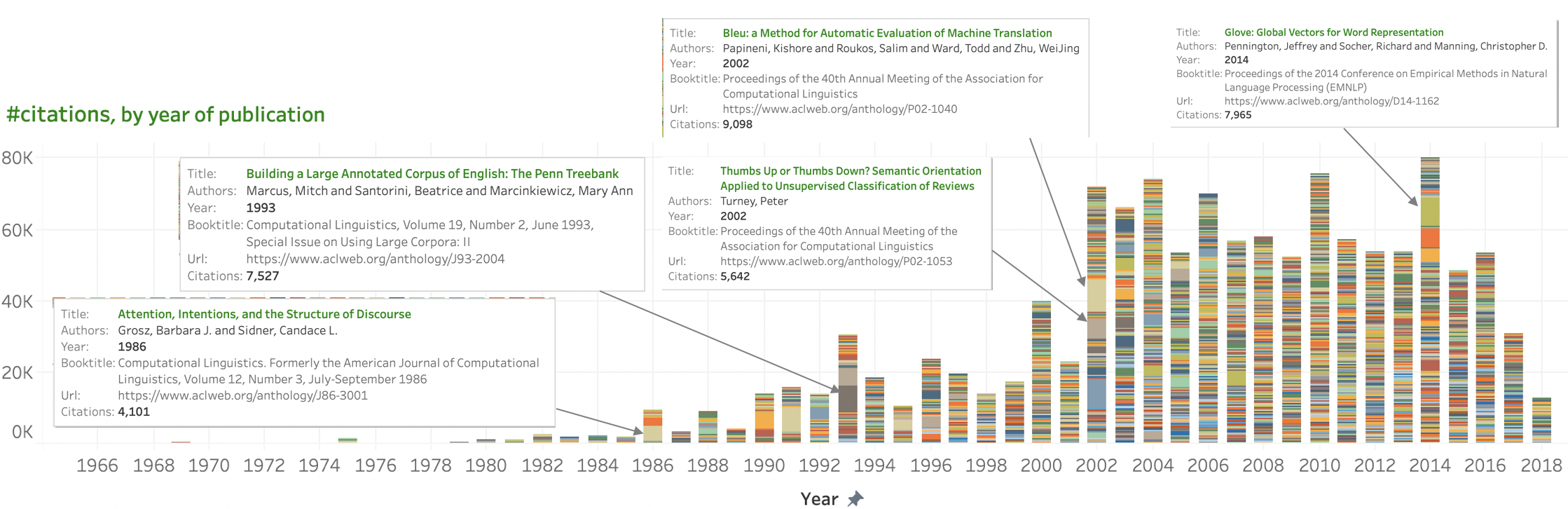}
 	\caption{NLP Scholar: Hovering over individual papers in B2 pops open an  information box showing the paper title, authors, year of publication, publication venue, and \#citations.}
 	\label{fig:citnHover}
 \end{center}
\end{sidewaysfigure*}


\begin{figure*}[t!]
 \begin{center}
 	\includegraphics[width=2\columnwidth]{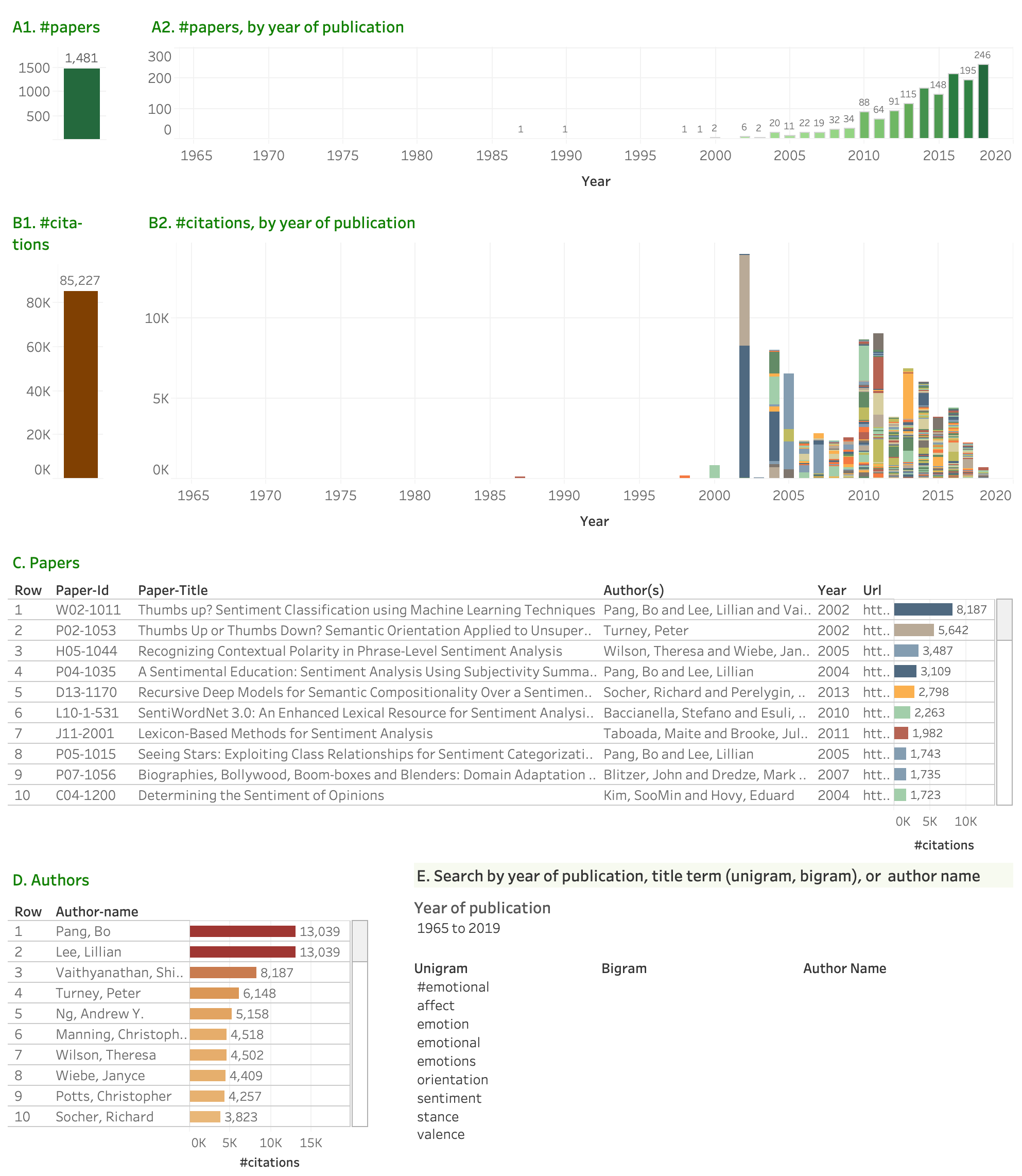}
 	\caption{NLP Scholar: After entering terms associated with sentiment analysis in the search box.}
 	\label{fig:SA}
 \end{center}
\end{figure*}


\begin{figure*}[t!]
 \begin{center}
 	\includegraphics[width=2\columnwidth]{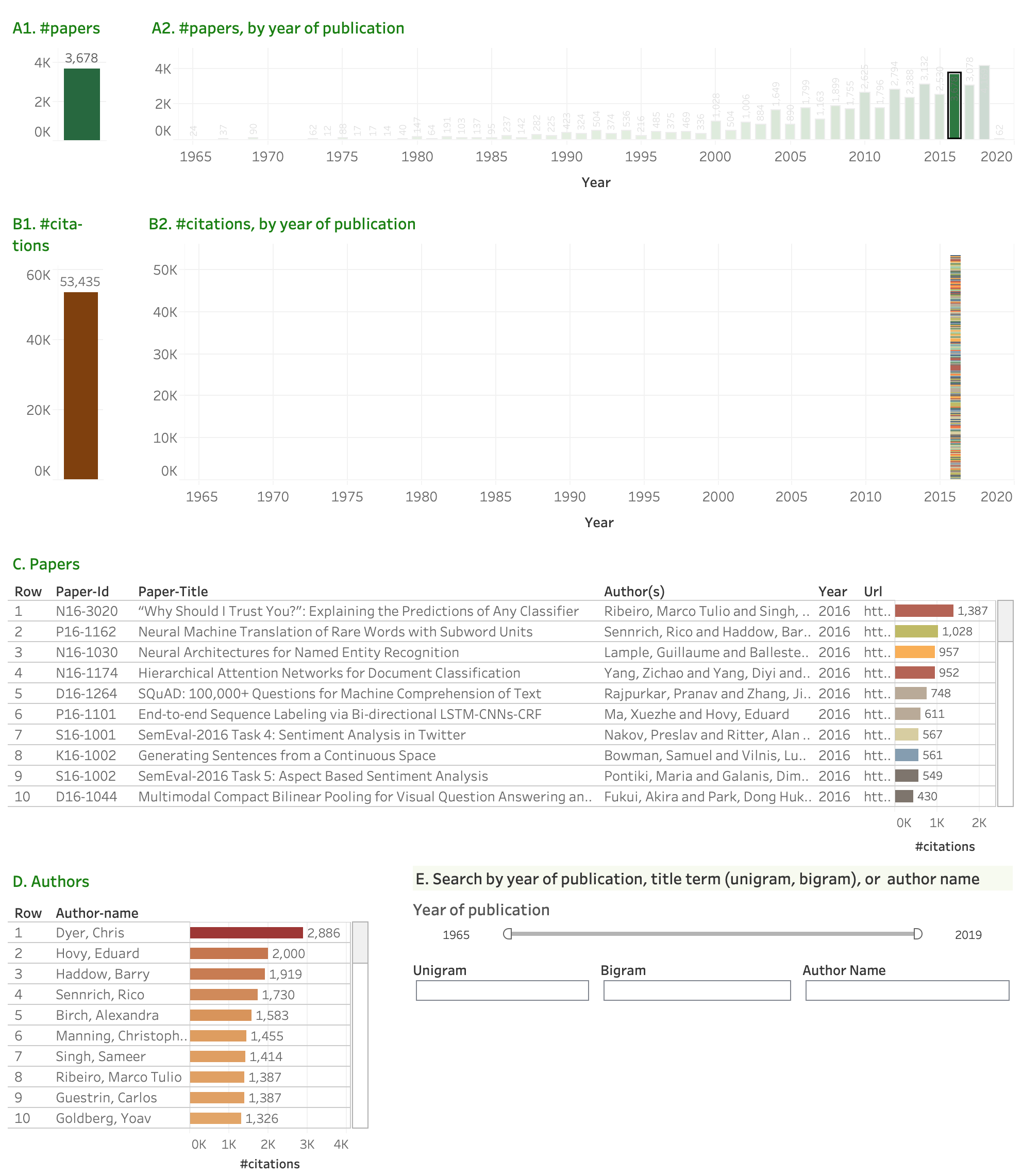}
 	\caption{NLP Scholar: After clicking on the 2016 bar in the \#papers by year viz (A2).}
 	\label{fig:2016}
 \end{center}
\end{figure*}


\begin{figure*}[t!]
 \begin{center}
 	\includegraphics[width=2\columnwidth]{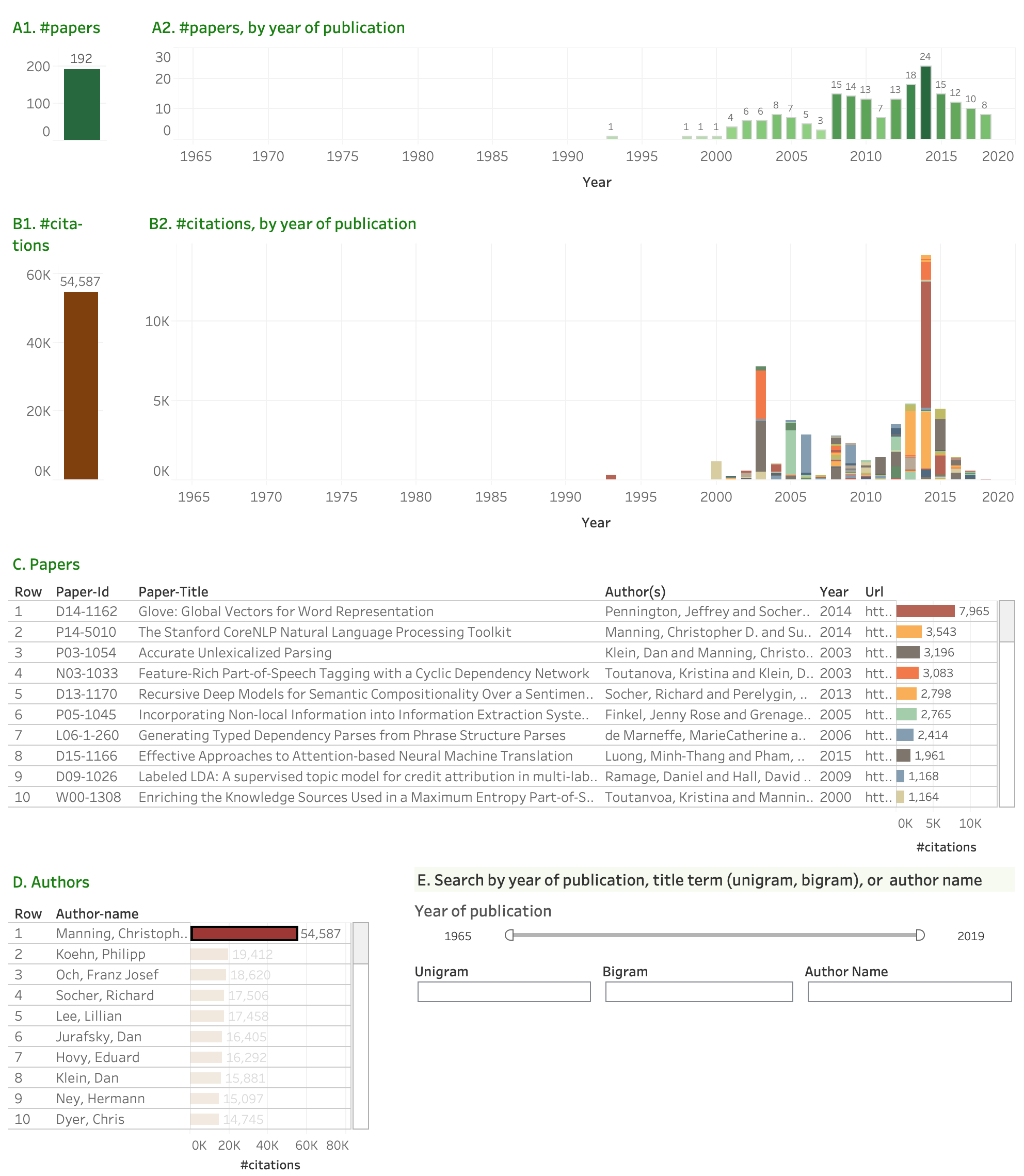}
 	\caption{NLP Scholar: After clicking on `Manning, Christopher' in the Authors list (D).}
 	\label{fig:Manning}
 \end{center}
\end{figure*}

\begin{figure*}[t!]
 \begin{center}
 	\includegraphics[width=2\columnwidth]{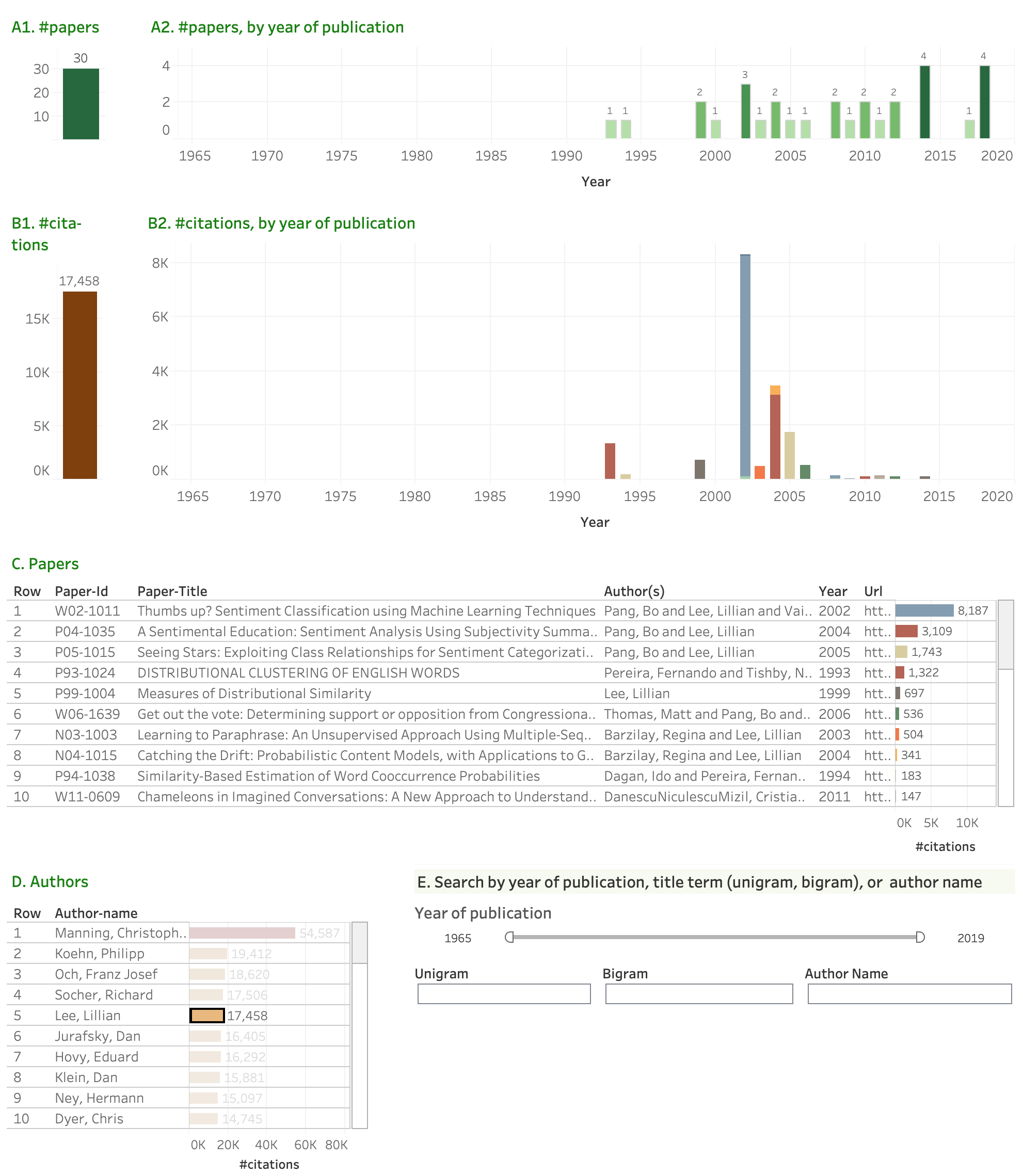}
 	\caption{NLP Scholar: After clicking on `Lee, Lillian' in the Authors list (D).}
 	\label{fig:Lee}
 \end{center}
\end{figure*}


\begin{figure*}[t!]
 \begin{center}
 	\includegraphics[width=2\columnwidth]{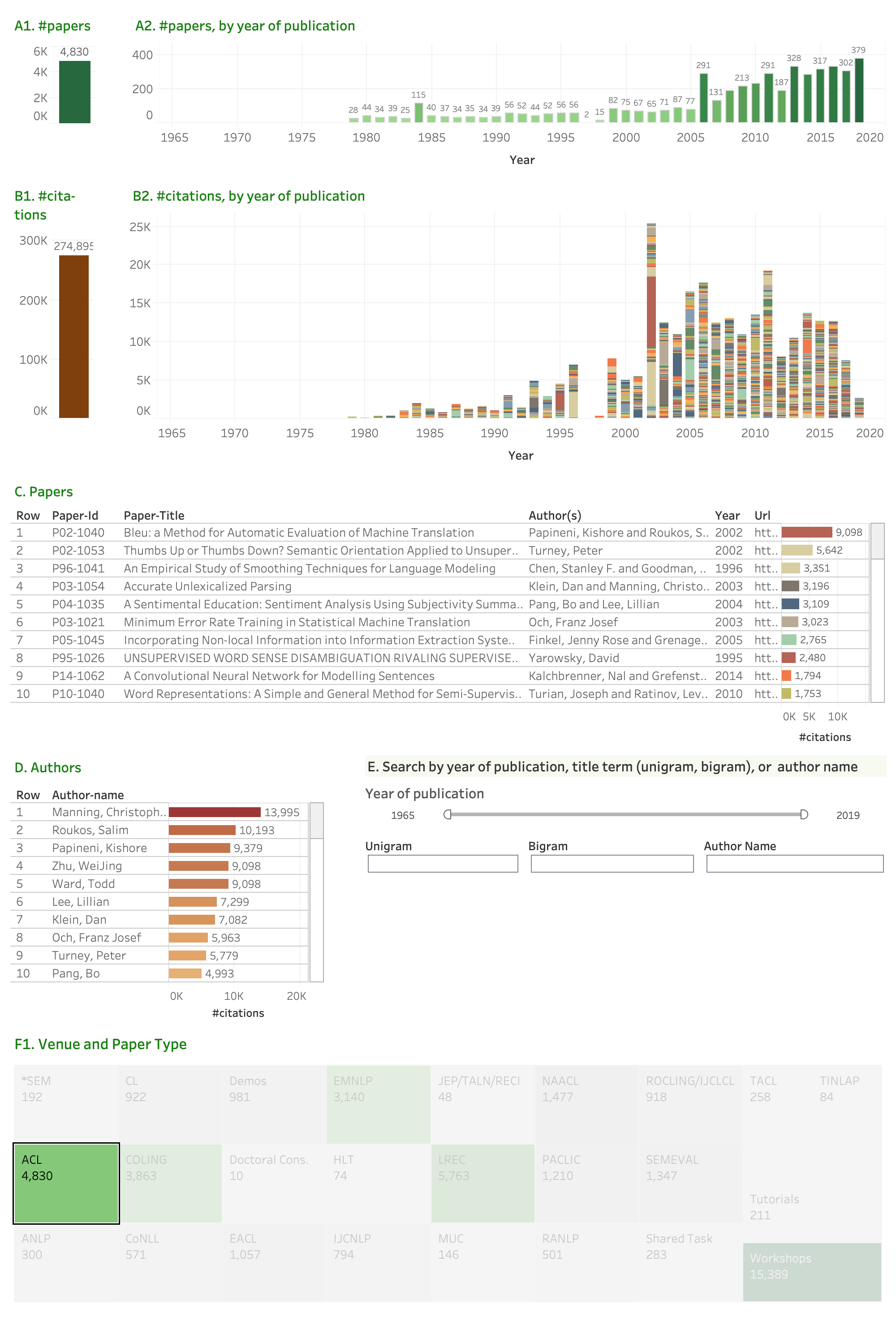}
 	\caption{NLP Scholar: After clicking on `ACL' in the venue and paper type treemap (F1).}
 	\label{fig:ACL}
 \end{center}
\end{figure*}

\begin{figure*}[t!]
 \begin{center}
 	\includegraphics[width=2\columnwidth]{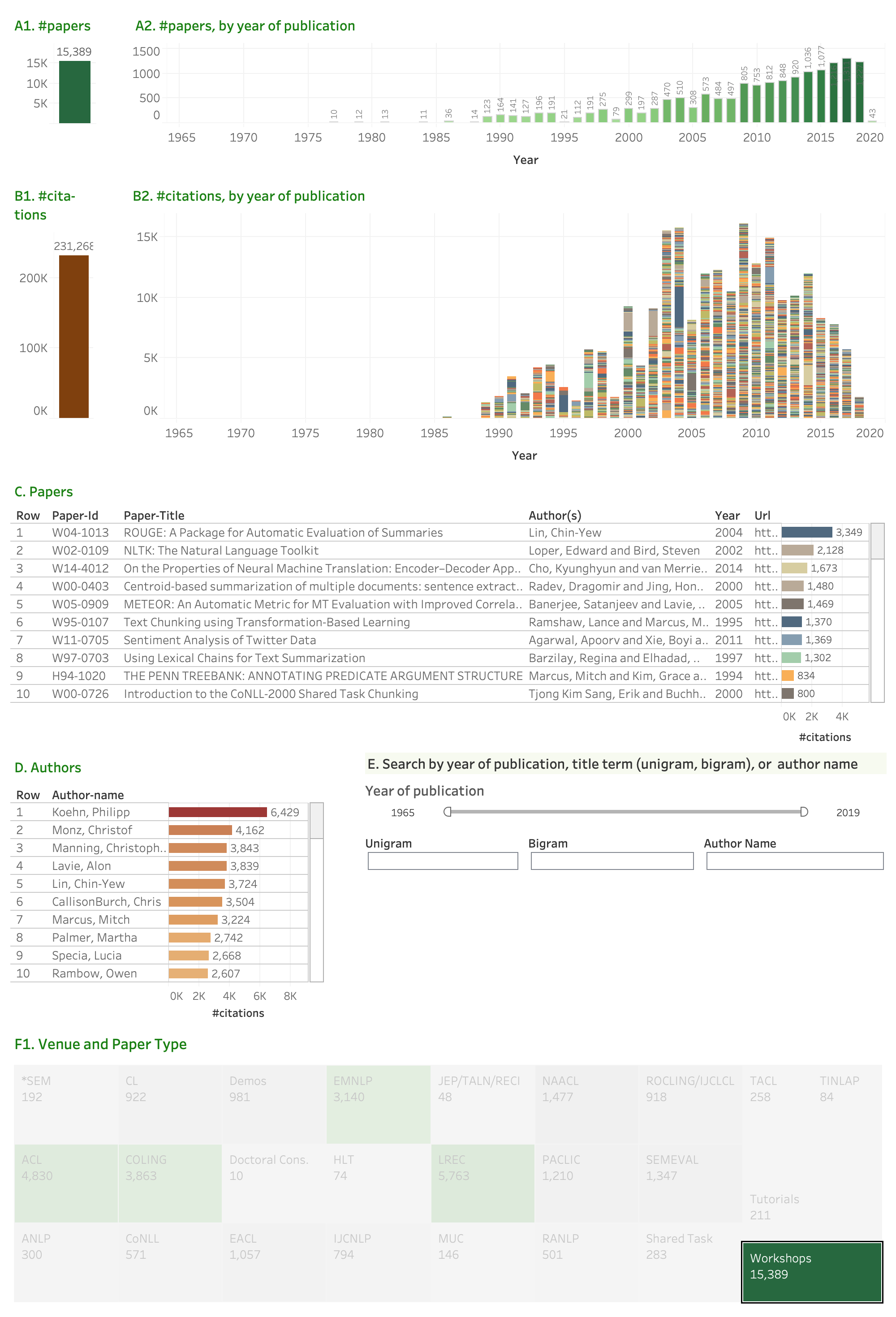}
 	\caption{NLP Scholar: After clicking on `Workshops' in the venue and paper type treemap (F1).}
 	\label{fig:workshops}
 \end{center}
\end{figure*}


\begin{figure*}[t!]
 \begin{center}
 	\includegraphics[width=2\columnwidth]{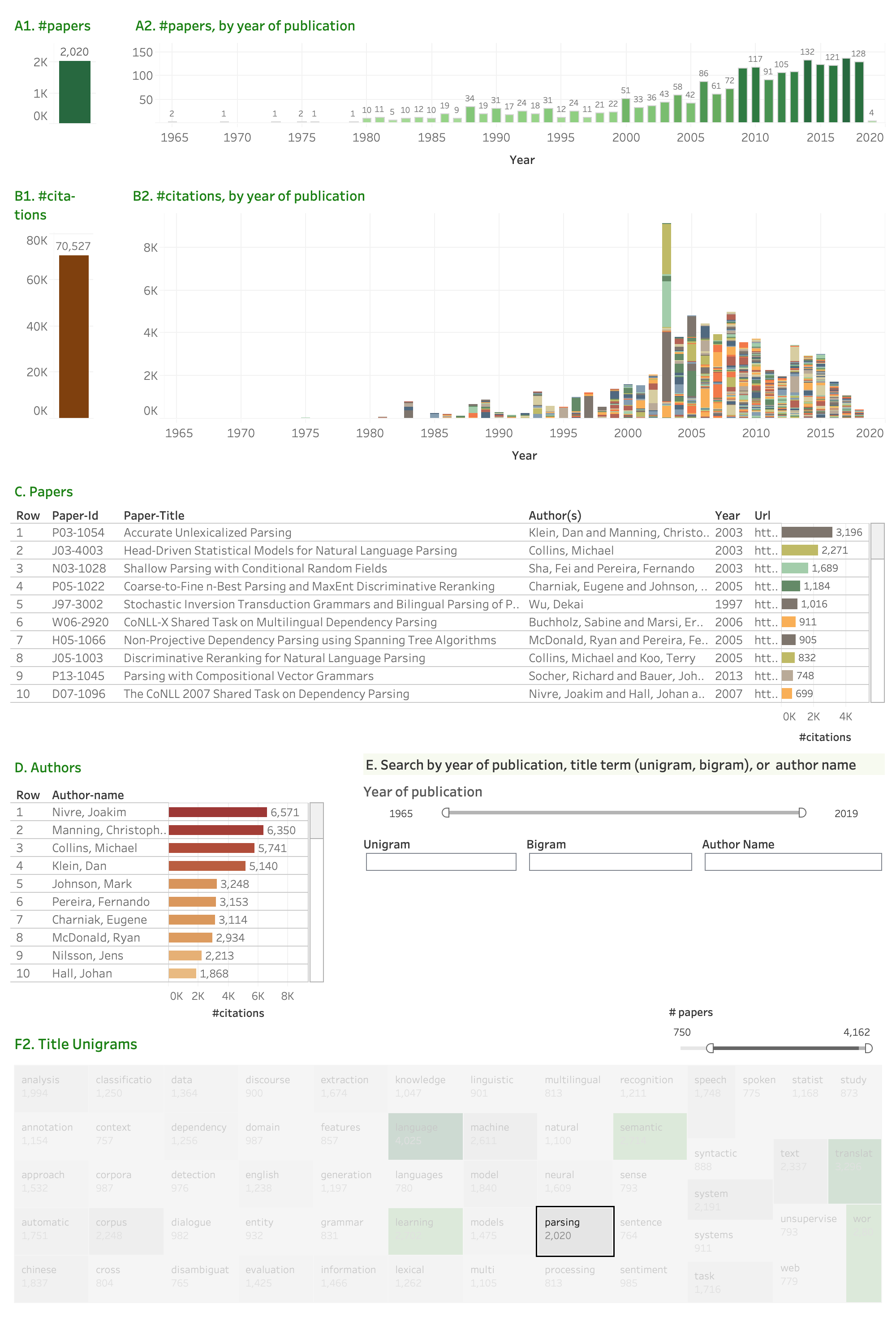}
 	\caption{NLP Scholar: After clicking on `parsing' in the unigrams treemap (F2).}
 	\label{fig:parsing}
 \end{center}
\end{figure*}

\begin{figure*}[t!]
 \begin{center}
 	\includegraphics[width=2\columnwidth]{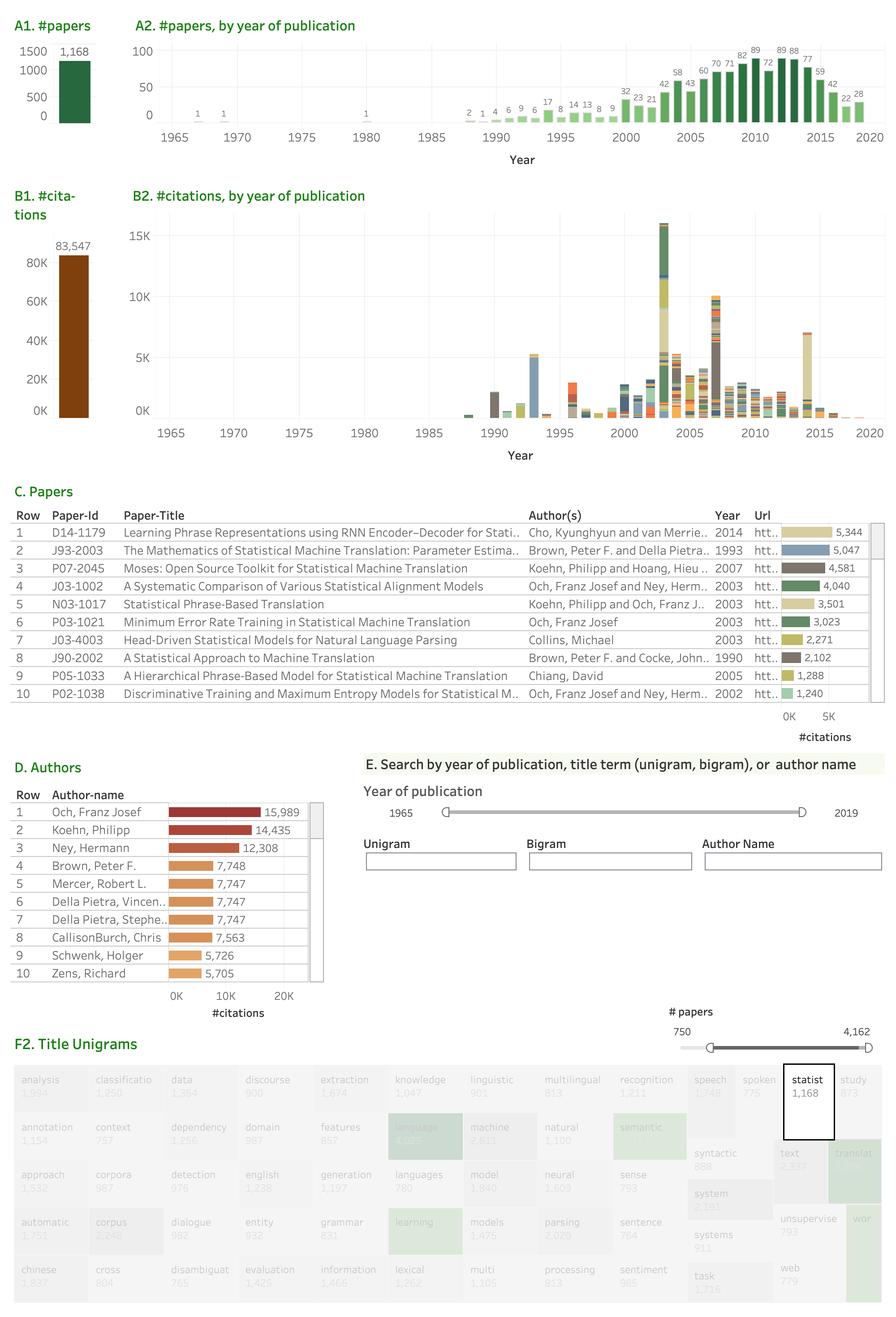}
 	\caption{NLP Scholar: After clicking on `statistical' in the unigrams treemap (F2).}
 	\label{fig:statistical}
 \end{center}
\end{figure*}

\begin{figure*}[t!]
 \begin{center}
 	\includegraphics[width=2\columnwidth]{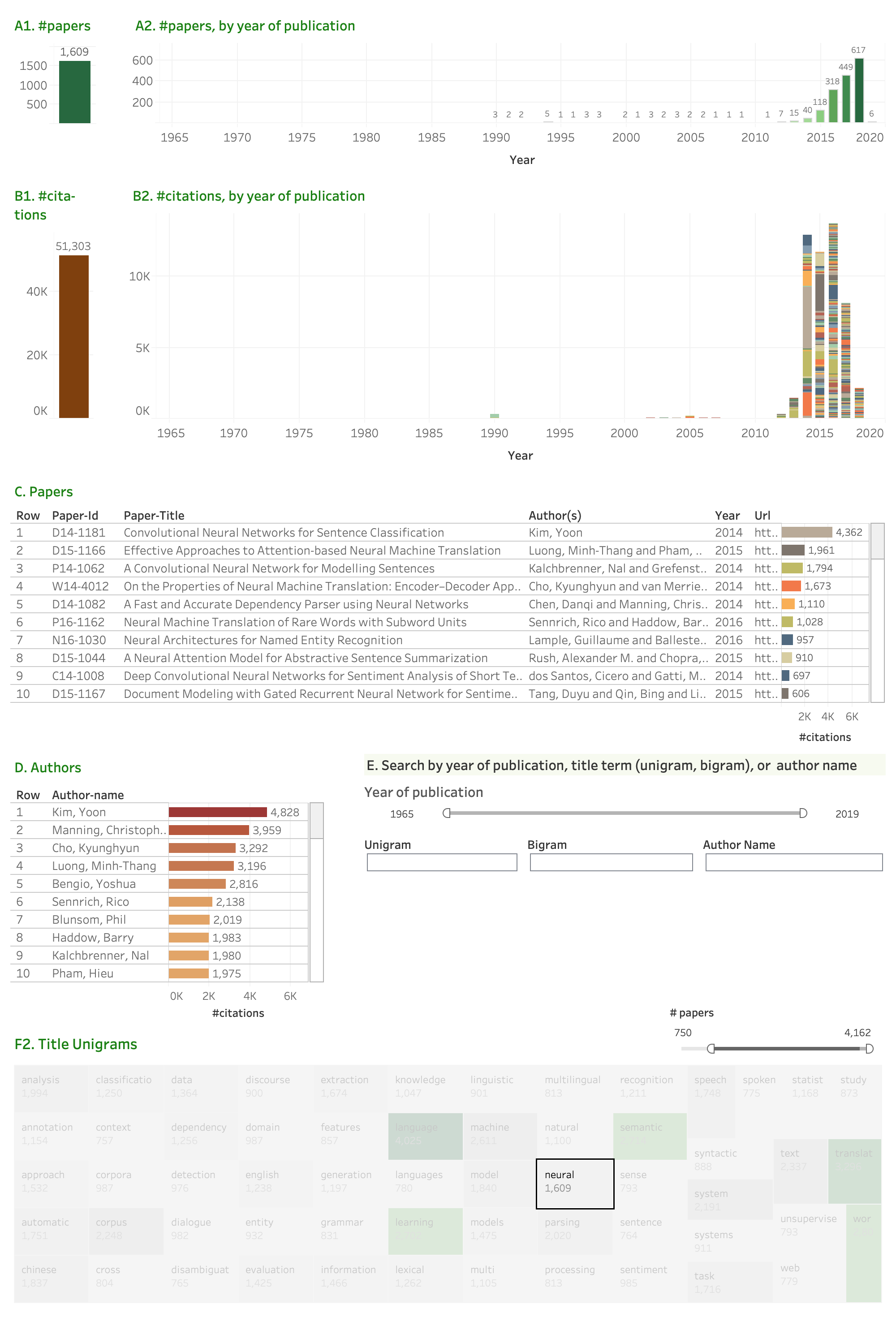}
 	\caption{NLP Scholar: After clicking on `neural' in the unigrams treemap (F2).}
 	\label{fig:neural}
 \end{center}
\end{figure*}


\begin{figure*}[t!]
 \begin{center}
 	\includegraphics[width=2\columnwidth]{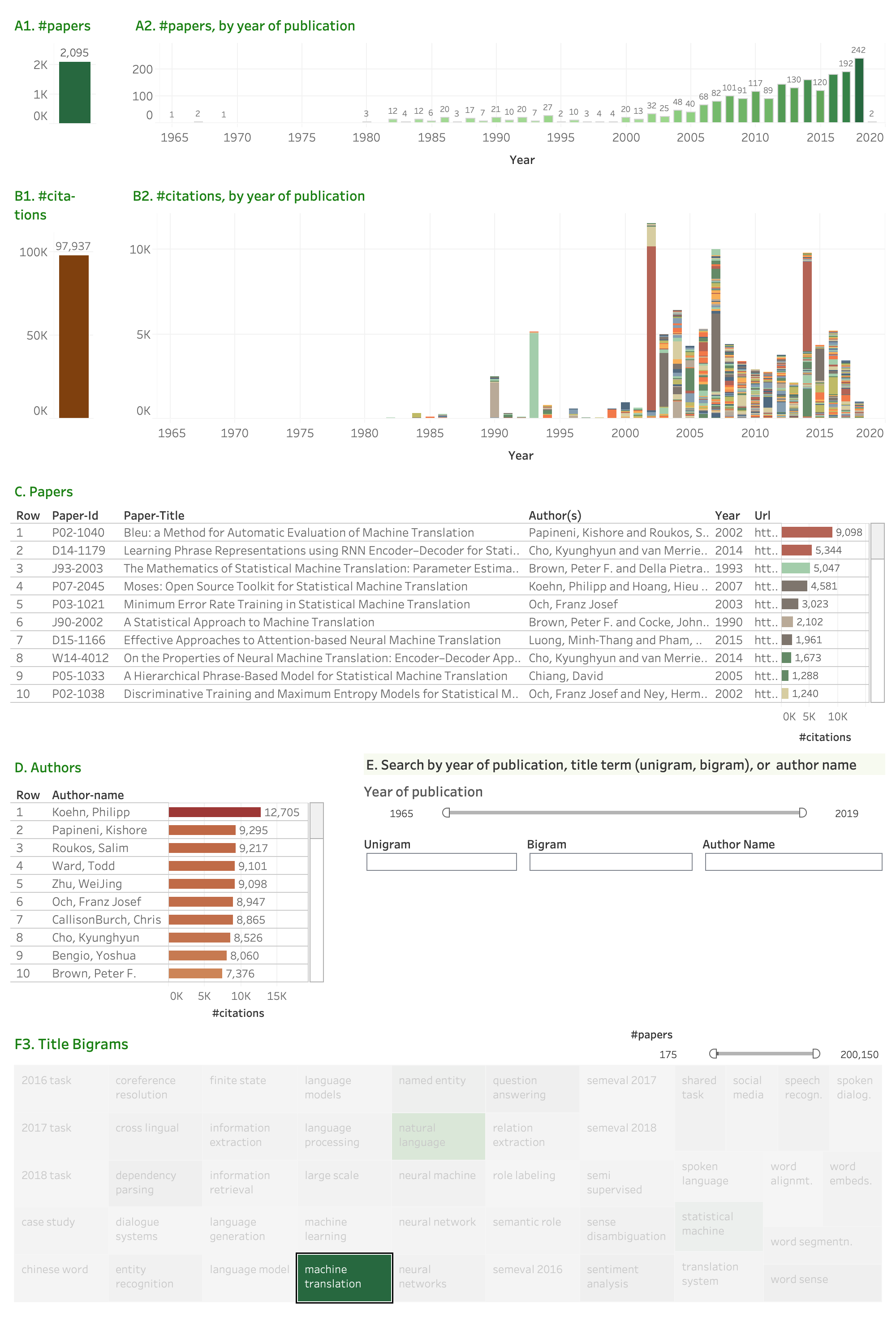}
 	\caption{NLP Scholar: After clicking on `machine translation' in the bigrams treemap (F3).}
 	\label{fig:MT}
 \end{center}
\end{figure*}

\begin{figure*}[t!]
 \begin{center}
 	\includegraphics[width=2\columnwidth]{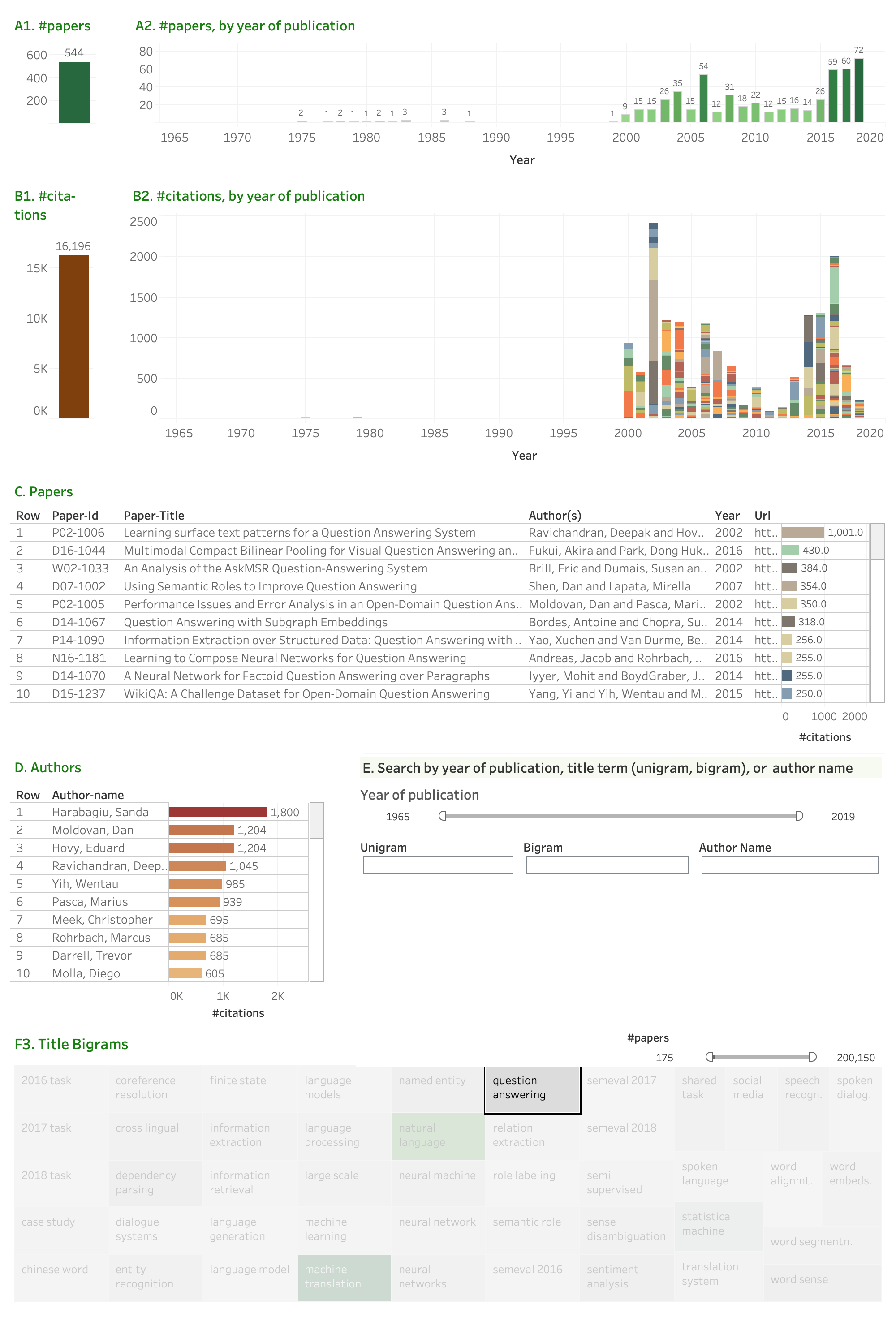}
 	\caption{NLP Scholar: After clicking on `question answering' in the bigrams treemap (F3).}
 	\label{fig:QA}
 \end{center}
\end{figure*}

\begin{figure*}[t!]
 \begin{center}
 	\includegraphics[width=2\columnwidth]{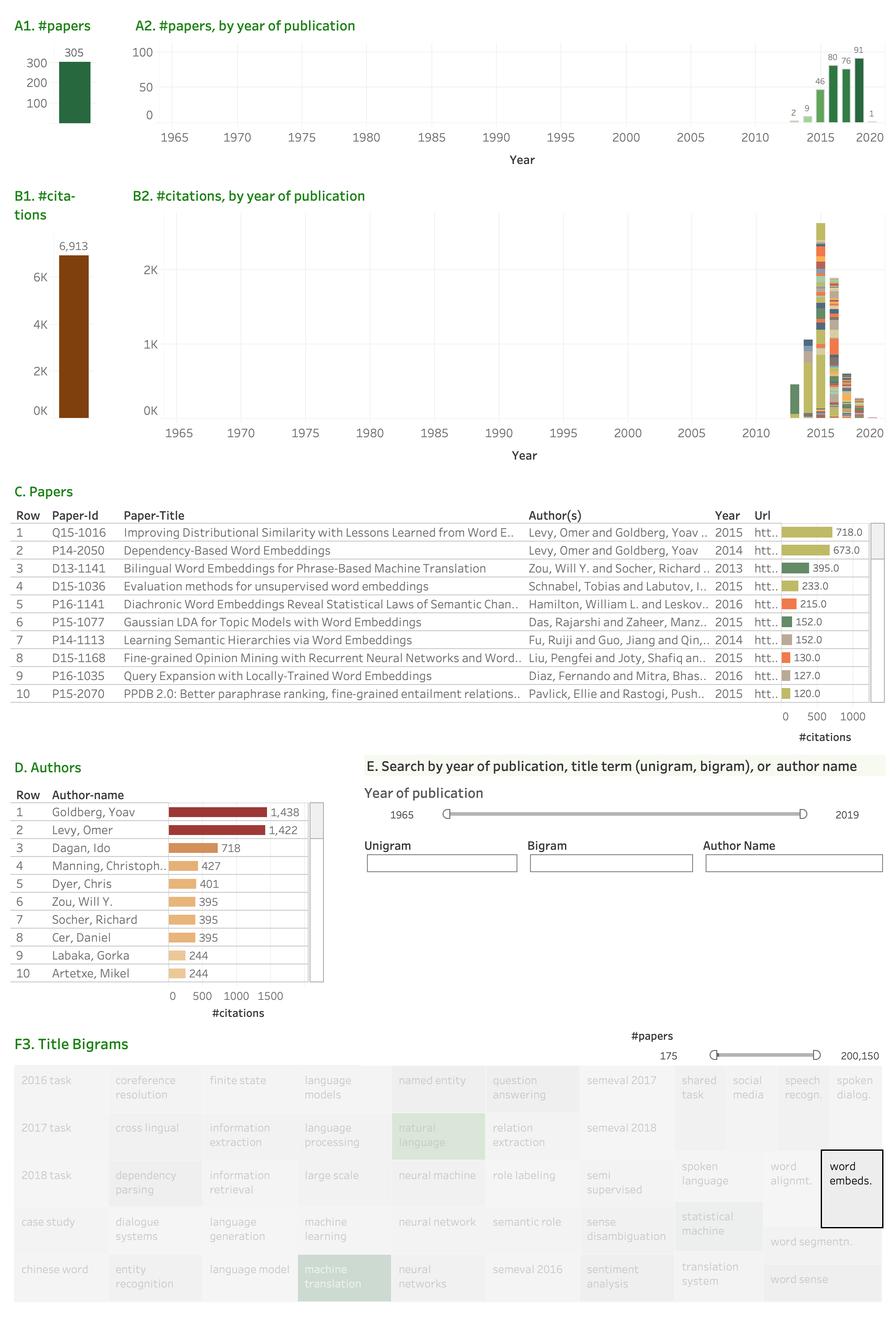}
 	\caption{NLP Scholar: After clicking on `word embeddings' in the bigrams treemap (F3).}
 	\label{fig:word-embeddings}
 \end{center}
\end{figure*}


\begin{figure*}[t!]
 \begin{center}
 	\includegraphics[width=2\columnwidth]{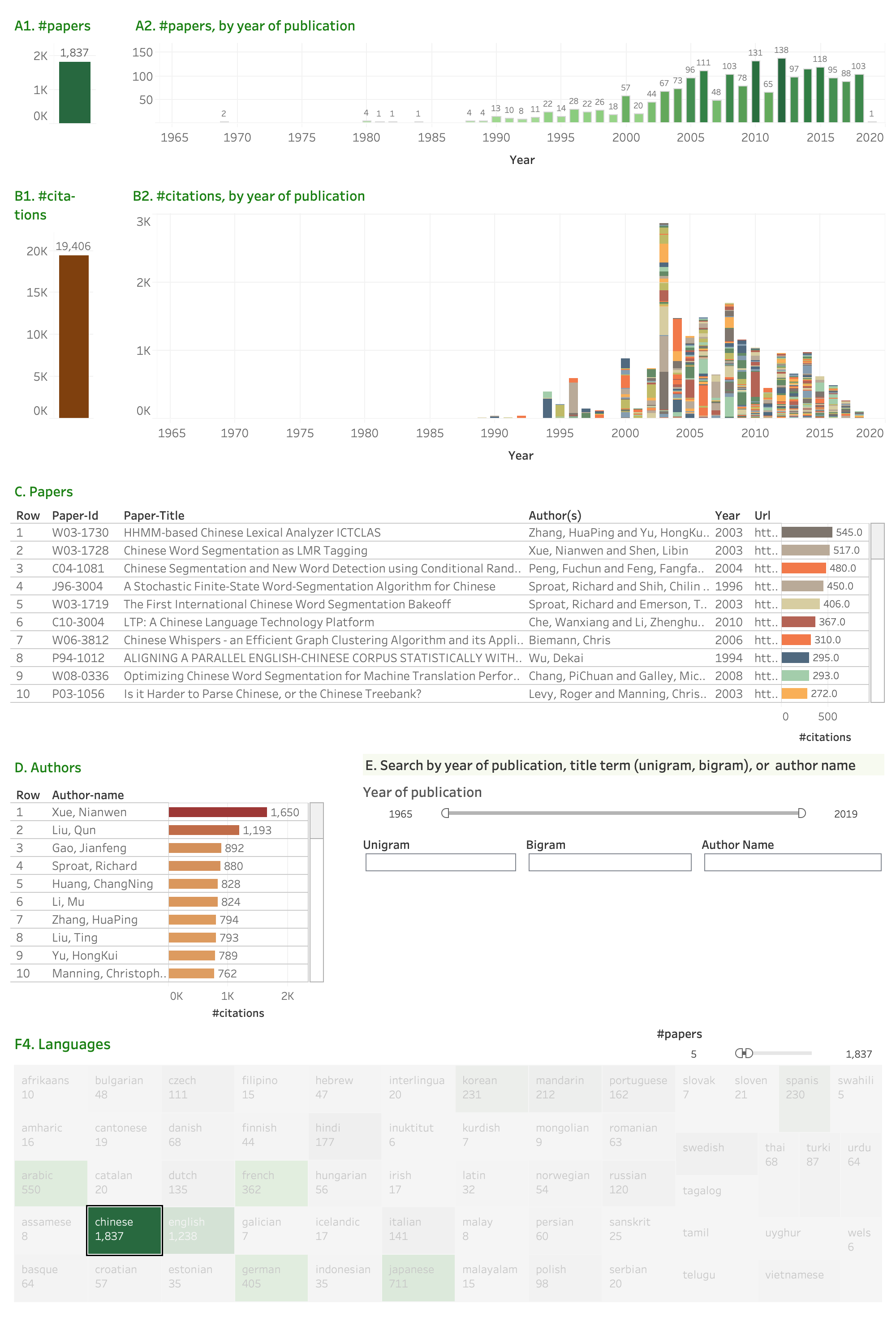}
 	\caption{NLP Scholar: After clicking on `Chinese' in the languages treemap (F4).}
 	\label{fig:Chinese}
 \end{center}
\end{figure*}

\begin{figure*}[t!]
 \begin{center}
 	\includegraphics[width=2\columnwidth]{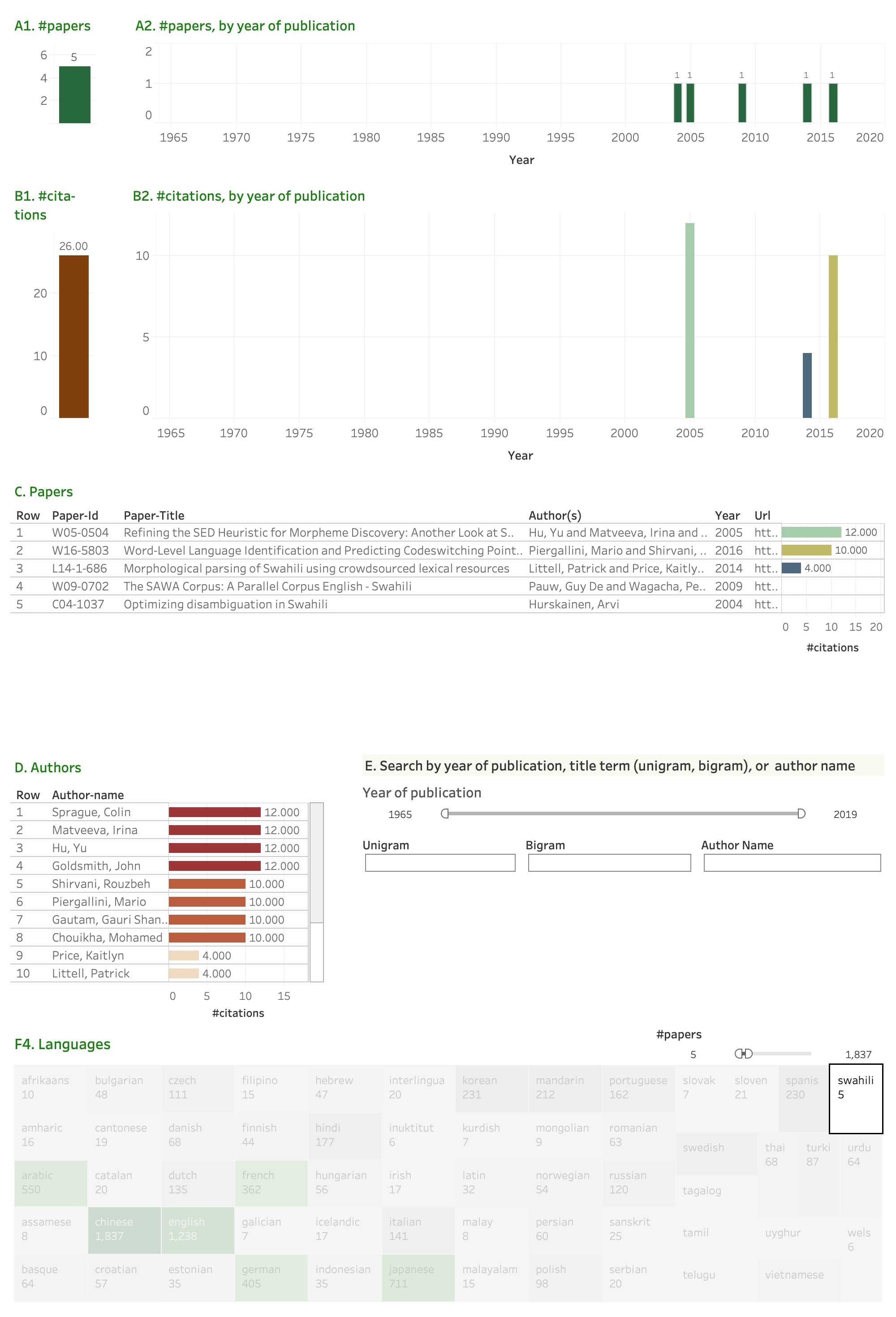}
 	\caption{NLP Scholar: After clicking on `Swahili' in the languages treemap (F4).}
 	\label{fig:Swahili}
 \end{center}
\end{figure*}


\end{document}